# Characterization of Fe-N nanocrystals and nitrogen–containing inclusions in (Ga,Fe)N thin films using transmission electron microscopy

A. Kovács,[1,a)] B. Schaffer,[2,3] M. S. Moreno,[4] , J.R. Jinschek[5], A. J. Craven,[3] T. Dietl,[6,7] A. Bonanni[8] and R. E. Dunin-Borkowski[1]

[1] Ernst Ruska-Centre for Microscopy and Spectroscopy with Electrons and Peter Grünberg Institute, Forschungszentrum Jülich, 52425 Jülich, Germany

[2] SuperSTEM, STFC Daresbury Laboratories, Keckwick Lane, Warrington WA4 4AD, United Kingdom

[3] SUPA, School of Physics and Astronomy, University of Glasgow, Glasgow G12 8QQ, United Kingdom

[4] Centro Atómico Bariloche, 8400 San Carlos de Bariloche, Argentina

[5] FEI Company, Achtseweg Noord 5, 5651 GG Eindhoven, The Netherlands

[6] Institute of Physics, Polish Academy of Sciences, al. Lotników 32/46, 02-668 Warszawa, Poland

[7] Institute of Theoretical Physics, Faculty of Physics, University of Warsaw, 00-681 Warszawa, Poland

[8] Institut für Halbleiter- und Festkörperphysik, Johannes Kepler University, Altenbergerstr. 69, 4040 Linz, Austria

a) Author to whom correspondence should be addressed. Electronic mail: a.kovacs@fz-juelich.de

**Abstract**

Nanometric inclusions filled with nitrogen, located adjacent to $Fe_nN$ ($n$ = 3 or 4) nanocrystals within (Ga,Fe)N layers, are identified and characterized using scanning transmission electron microscopy (STEM) and electron energy-loss spectroscopy (EELS). High-resolution STEM images reveal a truncation of the Fe-N nanocrystals at their boundaries with the nitrogen-containing inclusion. A controlled electron beam hole drilling experiment is used to release nitrogen gas from an inclusion *in situ* in the electron microscope. The density of nitrogen in an individual inclusion is measured to be 1.4 ± 0.3 g/cm$^3$. These observations provide an explanation for the location of surplus nitrogen in the (Ga,Fe)N layers, which is liberated by the nucleation of $Fe_nN$ ($n$> 1) nanocrystals during growth.

## I. INTRODUCTION

Recent progress in nanocharacterization[1] and *ab initio* studies[2,3] has shown that the open *d*-shells of transition metal (TM) cations diluted in non-magnetic compounds not only provide localized spins but also, through charge-state-dependent hybridization with band states, contribute to the cohesive energy of the material, particularly when TM atoms also occupy neighboring sites. The resulting attractive force between the magnetic cations may lead to their aggregation, either at the growth surface during the epitaxial process, as in (Ga,Fe)N (Refs. 4-7) and for Mn cation dimers in (Ga,Mn)As,[8] or by being triggered by appropriate post-growth high-temperature annealing[9-12] or high-temperature growth,[13] as observed in (Ga,Mn)As[9-12] and (Ga,In,Mn)As,[13] respectively. Significantly, in a number of systems, the TM-rich nanocrystals that are formed in this way, such as $Fe_nN$ ($n \geq 1$),[4-7] $MnAs$[13] or Co,[14-16] do not have a uniform distribution in the film. Instead, they tend to accumulate in planes that lie perpendicular to the growth direction, either close to the film surface[4-7,13] or at its interface with the substrate,[14-16] by a process that is referred to as nucleation-controlled aggregation.[6,16] One of the consequences of TM aggregation is that high temperature ferromagnetism in many magnetically-doped semiconductors and oxides is now assigned to the presence of such aggregates.[1,2,17] According to other schools of thought, defects[15,18] and electron-mediated interactions[19] account for robust ferromagnetism in some cases. Nanocomposite systems that contain ferromagnetic aggregates can also show enhanced magneto-optical[11] and magneto-transport properties,[20] including specific tunneling magnetoresistance.[21] A number of other functionalities are expected to be revealed in the future.[22,23]

Here, we make use of recent advances in aberration-corrected scanning transmission electron microscopy (STEM) and optimized specimen preparation techniques for electron microscopy to study, with high spatial resolution, (Ga,Fe)N layers that contain $Fe_nN$ nanocrystals, for which $n$ = 3 or 4. We use annular dark-field (ADF) imaging in the STEM to record images with atomic number sensitivity ($Z$ contrast). We show that the Fe$_n$N nanocrystals that form in the (Ga,Fe)N host are often truncated and are then associated with closely-adjacent inclusions that are filled with nitrogen. We use a combination of ADF STEM imaging and electron energy-loss spectroscopy (EELS) in an attempt to determine the nitrogen density in an individual inclusion. We also release the nitrogen from an inclusion *in situ* in the transmission electron microscope using a focused electron beam. Our results provide new information about the location of the nitrogen that is liberated from (Ga,Fe)N during the nucleation of Fe$_n$N ($n >$ 1) nanocrystals and have implications for understanding the physical properties of (Ga,Fe)N and other nanocomposite systems, such as GaAs/MnAs and (Zn,Co)O/Co.

**II. EXPERIMENTAL DETAILS**

(Ga,Fe)N samples were grown using metal organic vapor phase epitaxy on c-plane oriented sapphire substrates. TMGa, NH$_3$ and Cp2Fe were used as precursors for Ga, N and Fe, respectively, while H$_2$ was used as a carrier gas. The growth process was carried out as follows: substrate nitridation, low temperature deposition of a GaN nucleation layer that was annealed in the presence of NH$_3$ until recrystallization, followed by the growth of ~ 1 μm of a high-quality GaN at 1030 °C. Fe-doped GaN layers were deposited on the GaN buffer at temperatures ranging from 800 to 1050 °C. The deposition process, the structure of the layers and their magnetic properties are described in detail elsewhere.[24] (Ga,Fe)N layers that were grown at 800 °C showed no evidence of secondary phases. Here, we focus on layers that were grown either at 850 °C or at higher temperatures and contain Fe-N precipitates.

Structural characterization and chemical analysis were performed on cross-sectional specimens that had been prepared for TEM examination using conventional mechanical polishing and Ar ion milling. The procedure involved gluing a (Ga,Fe)N/sapphire sample to a Si single crystal using Gatan G1 glue. This structure was polished from both sides to a thickness of ~50 μm using diamond lapping paper with grain sizes of 30, 3 and 1 μm. A high-energy (3.5 kV) Ar ion beam was applied from the Si side while oscillating the specimen during ion milling. The ion energy was decreased progressively to 1 kV, while the reduction in specimen thickness was monitored by following the color change of the Si crystal optically in transmission. After perforation of the specimen, lower energy Ar ion milling at 0.5 kV from the specimen side was used to reduce surface damage.

Probe-aberration-corrected STEM studies were carried out at 300 kV and 100 kV using FEI Titan 80-300 and Nion UltraSTEM microscopes, respectively, with aberration functions corrected up to fourth order. The inner semi-angle of the ADF detector was varied between 24 and 78.4 mrad when collecting low-angle ADF (LAADF) and high-angle ADF (HAADF) signals. The STEM probe convergence and effective collection semi-angles used for EELS were both ~ 25 mrad in the experiments performed using the Titan microscope. For the dedicated EELS experiments carried out using the Nion microscope, the probe convergence and collection semi-angles were 30 and 33 mrad, respectively. EELS signals from molecular nitrogen gas alone were collected at room temperature at a nitrogen pressure of 20 mbar using an FEI Titan 80-300 environmental TEM (ETEM) operated at 300 kV. N-*K* edge EELS fine structures in GaN were calculated using self-consistent real-space multiple-scattering calculations[25] implemented in FEFF9.05 density functional theory code, which allows experimental parameters such as electron beam energy, crystal orientation and collection angle to be included. The random phase approximation was used to include core hole effects, while the Hedin-Lundqvist self-energy was used to take inelastic losses into account.

The crystallographic structures of the Fe-N nanocrystals were determined by using a highly parallel electron beam with full-width at half maximum of ~ 1 nm to record nano-beam electron diffraction (NBED) patterns, which were compared with simulated patterns were generated using JEMS software.

## III. RESULTS

### A. Structural analysis

A representative low magnification LAADF STEM image of a (Ga,Fe)N layer that had been grown at 900 °C is shown in Fig. 1 (a). Both dislocations and Fe-N nanocrystals appear bright in the image. The dark contrast adjacent to each nanocrystal, which we observed in every (Ga,Fe)N sample that contained Fe-N nanocrystals larger than ~5 nm, is an inclusion filled with nitrogen, as discussed below. The structures of the nanocrystals were determined, using NBED (see below), to be $\varepsilon$-$Fe_3N$ or $\gamma$-$Fe_4N$, in agreement with previous diffraction and magnetization measurements.[24] High-resolution aberration-corrected ADF STEM images of an individual nanocrystal and an adjacent nitrogen inclusion recorded using different inner detector semi-angles are shown in Figs. 1 (b) and (c). The dissimilar crystallographic structures of the Fe-N nanocrystal and the surrounding GaN matrix result in the formation of a Moiré fringe pattern within the outline of the crystal in Fig. 1(b). The image shows that the nanocrystal is faceted, with a truncated hexagonal shape, as marked in Fig. 1(b). The volume of the missing part of the crystal is ~ 32% of the volume that it would have had if it were not truncated. By considering a nanocrystal that has the structure and composition of $\varepsilon$-$Fe_3N$ and molecular nitrogen, the nitrogen content of the missing part of the nanocrystal is equivalent to the volume of a ~ 6 nm nitrogen-filled bubble at room temperature and pressure. The size of the inclusion shown in Fig. 1 (b) is, however, larger than 10 nm, suggesting either that excess nitrogen may have been released during nucleation of the nanocrystal or that the inclusion contains nitrogen at a different pressure. The thin bright band of contrast that is visible around the inclusion in the LAADF image shown in Fig 1 (b) may be associated with strain[26] and depends sensitively on collection angle and sample thickness. Significant segregation of Fe, N or Ga was ruled out as an explanation for the origin of the contrast by acquiring EELS line scans across the edge of the inclusion. By increasing the collection angle of the detector to acquire HAADF image, as shown in Fig. 1(c), the contrast is more sensitive to projected atomic number density and less to diffraction contrast. The inclusion then appears with dark contrast in the recorded HAADF image.

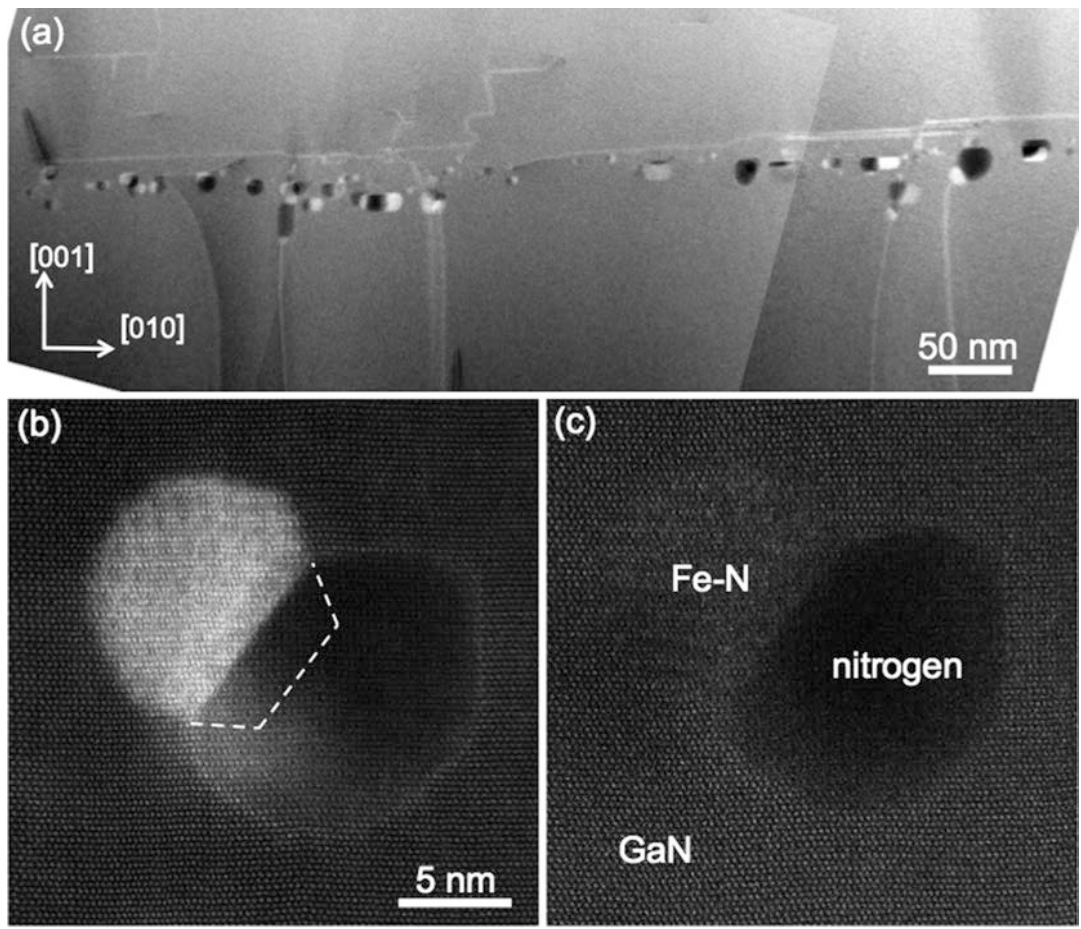

**Figure 1**. (a) Montage of low magnification LAADF STEM images of Fe-N nanocrystals and nitrogen-containing inclusions in a GaN layer that had been grown at 900 °C. (b) LAADF and (c) HAADF images of a 10 nm Fe-N nanocrystal and an associated nitrogen inclusion. The region indicated in (b) shows an apparently truncated part of the crystal. The inner ADF detector semi-angles used were (a) 47.4, (b) 30.9, and (c) 78.4 mrad, respectively.

In each sample, Fe-N nanocrystals with a size of ~ 5 nm were also found without nitrogen-containing inclusion adjacent to them. Figure 2 (a) shows an aberration-corrected high-resolution LAADF STEM image of a 4.5 x 3 nm Fe-N nanocrystal in a sample that had been grown at 950 °C. A Moiré fringe pattern is visible across the nanocrystal due to the overlapping Fe-N and GaN structures. A different nanocrystal from the same sample was studied using NBED as shown in Fig. 2 (b). A diffraction patterns were recorded both from the Fe-N nanocrystal and from the GaN matrix, which was used as a standard for lattice parameter determination. This procedure was used to establish that the nanocrystal was ε-$Fe_3N$. Figure 2 (c) shows simulated diffraction pattern of ε-$Fe_3N$ and GaN, which provide a good qualitative match to the experimental pattern shown in Fig. 2 (b). The epitaxial relationship is inferred to be (001)[100]GaN // (001) [210]ε-$Fe_3N$. The simulated diffraction pattern was determined using lattice parameters for ε-$Fe_xN_y$ obtained from Leineweber et al.[27] The lattice parameter of the ε-$Fe_3N$ nanocrystal, measured experimentally along the *b* axis, is 0.455±0.01 nm, which is slightly shorter than that of the bulk ε-phase with a composition of ε-$Fe_3N$, which is 0.469 nm. Such a lattice distortion can be caused either by strain or by a non-stoichiometric nanocrystal composition. The results of a compositional measurement across an ε-$Fe_3N$ nanocrystal and the GaN host, made by collecting a line-scan of N-*K* edge and Fe-*L* edge intensities from EELS spectra, are shown in Figs. 2 (d) and (e). A small dip in the measured N concentration and a clear Fe peak are consistent with the presence of an Fe-rich nanocrystal.

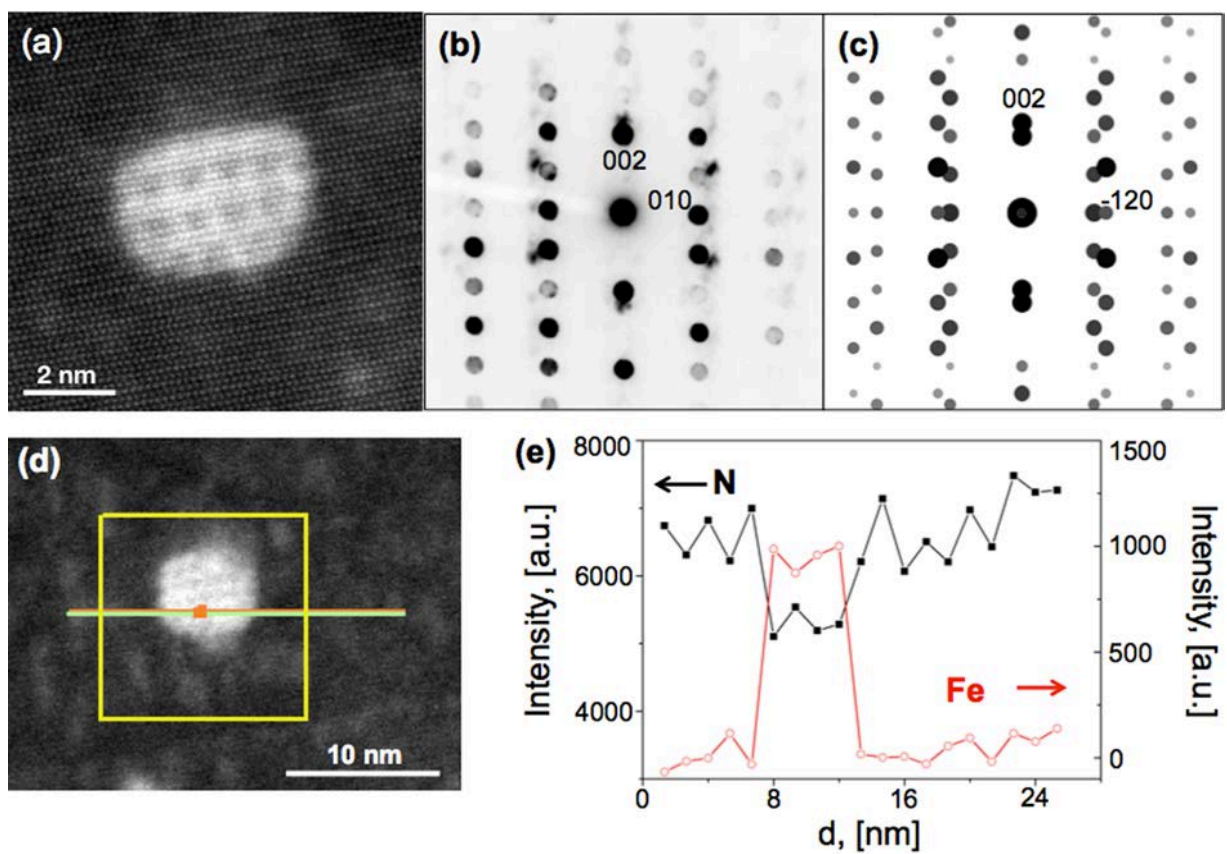

**Figure 2.** (a) LAADF STEM images and diffraction patterns acquired at 300 kV from ε-Fe$_3$N nanocrystals without adjacent nitrogen inclusions. (b) Experimental and (c) simulated NBED patterns of an ε-Fe$_3$N nanocrystal in a GaN host. (d) LAADF STEM image of a different crystal, showing the region that was used for subsequent EELS analysis. (e) EELS intensities corresponding to Fe (red) and N (black) signals recorded along the line indicated in (d).

High-resolution HAADF STEM images and diffraction patterns acquired from a γ-Fe$_4$N nanocrystal in GaN are shown in Fig. 3 for a sample that had been deposited at 950 °C. The nanocrystal had dimensions of approximately 50 x 26 nm. However, part of it is missing, where a nitrogen-containing inclusion has formed, as shown in Fig. 3 (a). The relatively large size of the nanocrystal allowed a conventional selected area electron diffraction (SAED) pattern to be recorded, showing weak reflections from the nanocrystal in addition to the reflections from GaN. Figures 3 (b) and (c) show experimental and simulated SAED patterns, from which the epitaxial relationship was inferred to be $(002)[100]GaN \| (-111)[110]\gamma\text{-}Fe_4N$. Interestingly, a tetragonal distortion of -2.4 % was inferred in the measured lattice spacing of the γ-$Fe_4N$ nanocrystal using the GaN reflections as a reference. A high-resolution HAADF STEM image of the lower interface between the γ-$Fe_4N$ nanocrystal and the GaN, in which the bright dots are likely to correspond to Fe and Ga columns, is shown in Fig. 3 (d). The structure of this interface is particularly interesting, since it is incoherent, with no dislocations observed in the γ-$Fe_4N$ despite the misfit of 5.4 % between the $(111)_{Fe\text{-}N}$ and $(011)_{GaN}$ lattice plane spacing. Moreover, a gap of ~0.34 nm is present between the γ-$Fe_4N$ and GaN, as shown in Fig. 3 (e). The measured Ga-Ga peak-to-peak distance of 0.26±0.1 nm in GaN and the measured Fe-Fe distance of 0.215±0.1nm in γ-$Fe_4N$ are close to the values of 0.259 and 0.216 nm expected for these structures. In the [111] direction, the γ-$Fe_4N$ structure consists of modulated Fe and N layers. It is reasonable to suggest that the first layer of the γ-$Fe_4N$ nanocrystal is N-rich, based on the dark contrast visible in the gap in the HAADF STEM image of the interface. The schematic model shown in Fig. 3 (f) illustrates the possible interface structure. A high-resolution HAADF STEM image of the orthogonal interface between the γ-$Fe_4N$ nanocrystal and the GaN host is shown in Fig. 3 (g). The misfit between the $(002)_{GaN}$ and $(111)_{Fe\text{-}N}$ planes is 16.6 %, resulting in the presence of periodic dislocations in the γ-$Fe_4N$ nanocrystal, as indicated in Fig. 3 (f). The dislocations formed every 4-5 planes, with a distance of 0.9 nm to 1.1 nm between them.

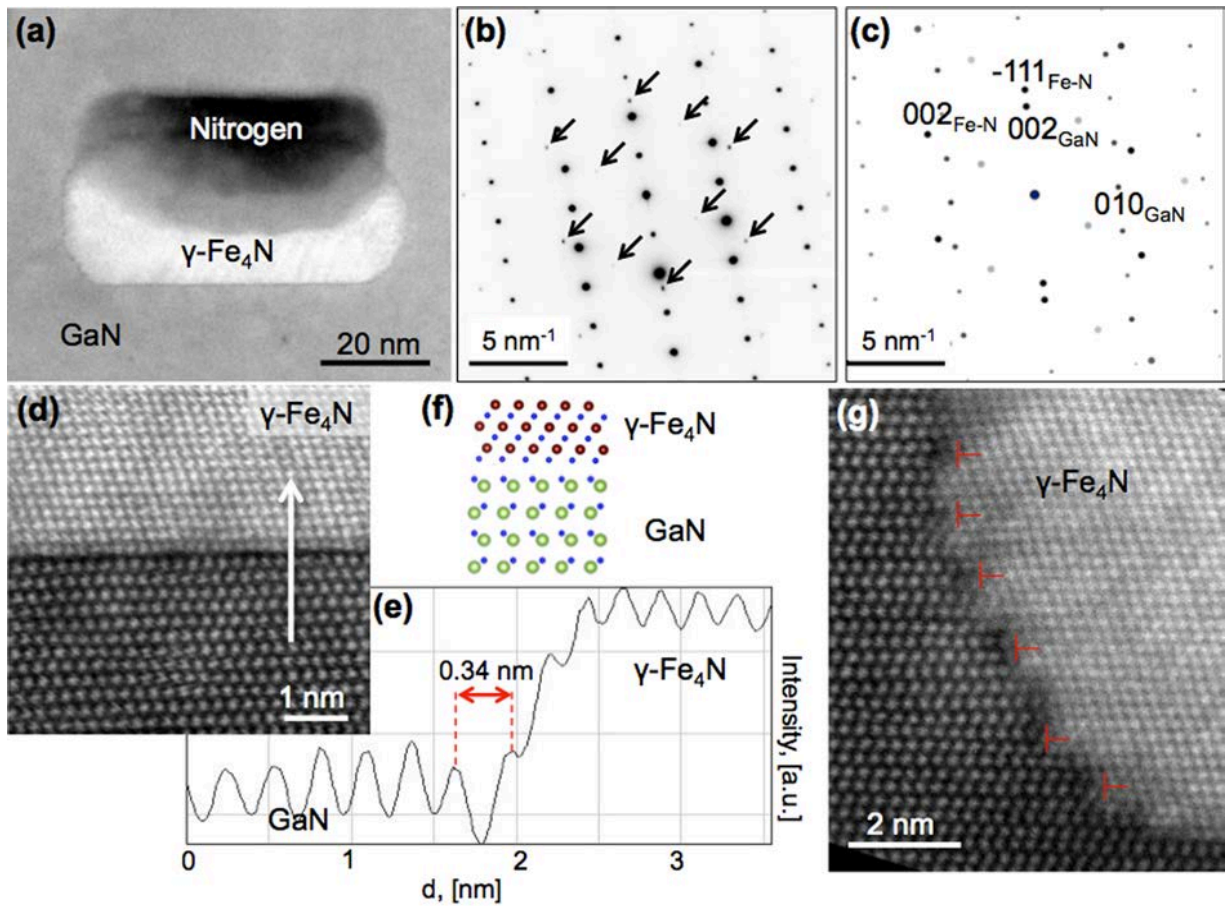

## B. EELS analysis of a nitrogen-containing inclusion

A STEM EELS measurement was performed to obtain chemical information from a single nanocrystal and an adjacent inclusion embedded in the GaN host. The measurement was challenging as a result of the presence of nitrogen in each of the three phases (GaN, Fe-N and nitrogen). We studied the fine structure of the N-$K$ edge by using a distributed dose acquisition routine (SMART[28]) to minimize electron-beam-induced damage during the experiment, which was performed at 100 kV. An ADF STEM image and background-subtracted EELS spectra acquired from an Fe-N nanocrystal and an associated inclusion in a (Ga,Fe)N layer that had been grown at 850 °C are shown in Fig. 4. The EELS line-scan spectra in Fig. 4 (b) were acquired from the area indicated by a box and an arrow shown in Fig. 4(a). Representative N-*K* edge spectra recorded from the GaN host, the (inclusion + GaN), and the (Fe-N nanocrystal + GaN) are shown in Fig. 4 (c). The spectrum recorded from the GaN host shows a characteristic three-peaked structure between 400 and 405 eV. This feature also appears in EELS spectra collected from the Fe-N/ inclusion complex, as they are embedded in the GaN host. However, the first peak in the spectrum that was collected from the inclusion, at 400 eV, is significantly higher than that recorded from either the Fe-N particle or the GaN alone. By normalizing the N-*K* edge tails, a difference in the heights of the second and third peaks of the N-*K* edge appears between the spectra recorded from the GaN and (GaN + inclusion) regions. This difference is associated with the contribution of the inclusion to the peak intensities. In order to interpret the fine structure of the N-*K* edge spectra, an EELS spectrum was recorded from molecular $N_2$ gas alone in an environmental TEM.[29] A characteristic single-peaked feature in the experimental spectrum recorded from nitrogen gas and multiple peaks in the spectrum simulated for GaN are visible in Fig. 4 (d). Distinct peaks in the experimental molecular nitrogen spectrum at ~ 415 eV and in the simulated GaN spectrum at ~ 423 eV can also be seen in the experimental spectra shown in Fig. 4 (c), suggesting that the spectrum recorded from the (inclusion + GaN) is indeed a superposition of spectra from molecular nitrogen and GaN.

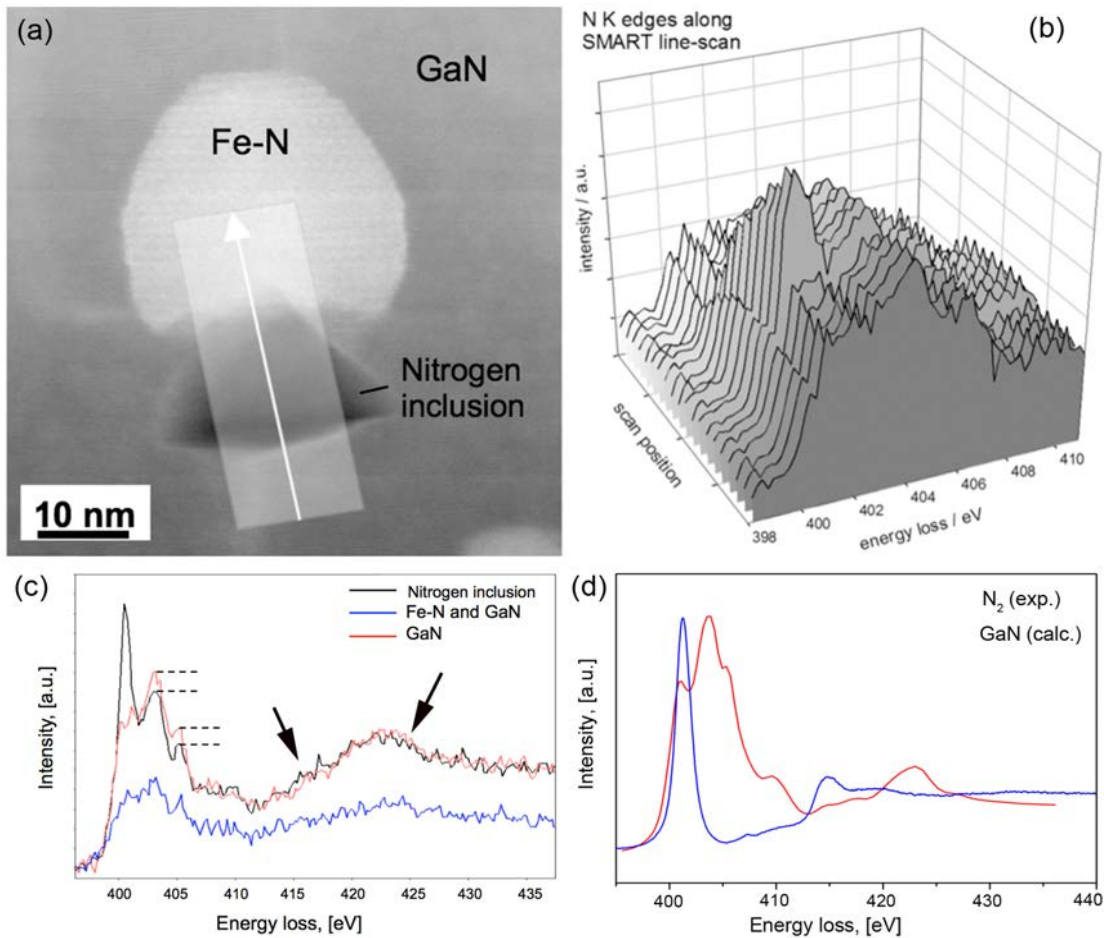

**Figure 4.** (a) ADF STEM image and (b) background-subtracted N-*K* edge EELS spectra acquired at 100 kV from an inclusion, an Fe-N nanocrystal and the GaN host. The inner ADF detector semi-angle used was 52 mrad. The box and arrow in (a) show the positions used for the line-scan measurements. (c) Representative EELS spectra recorded from the (inclusion + GaN), (GaN + Fe-N crystal) and GaN. The dotted lines in (c) indicate a difference in amplitude associated with the reduced GaN thickness at the position of the (inclusion + GaN). The arrows indicate distinct peaks associated with nitrogen and GaN (see (d)). (d) Experimental EELS spectrum recorded from nitrogen gas in an ETEM at 300 kV, shown alongside a spectrum calculated for the N-*K* edge in GaN.

Figure 5 shows the result of an experiment that provides direct evidence for the presence of nitrogen in the inclusion adjacent to the Fe-N nanocrystal shown in Fig. 4, obtained by making use of a focused electron beam to burst the inclusion *in situ* in the electron microscope. A stationary sub-Å-diameter electron beam with a current of 350 pA was used to create a hole in the specimen at the position of the inclusion, while recording an EELS spectrum every 40 s. Figures 5 (a) and (b) show ADF STEM images of the nitrogen inclusion and part of the adjacent Fe-N nanocrystal recorded before and after hole formation, respectively. The inclusion shape can be seen to change during the experiment. The intensity of the characteristic first peak in the N-*K* edge spectrum at 400 eV was observed to decrease suddenly when the nitrogen was released after irradiation for 600 s, as shown in Figs. 5 (c)-(e). After hole formation, the N-*K* edge fine structure is the same as that measured from GaN alone (see Figs. 4 (b) and (c)).

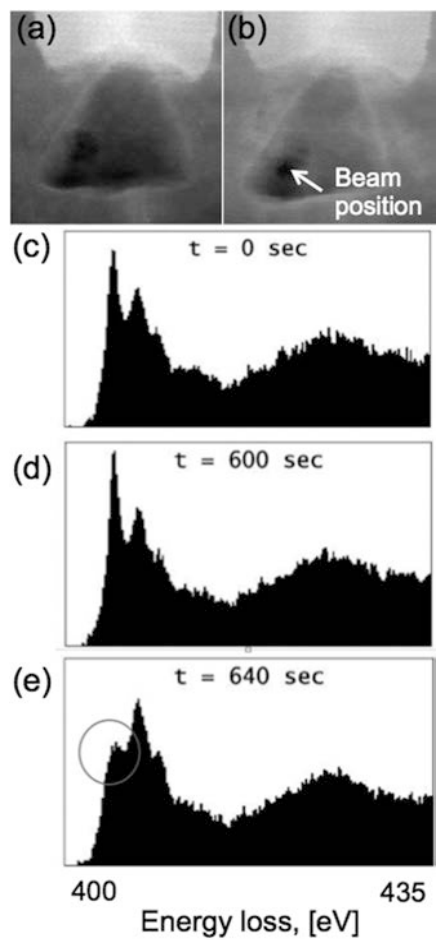
14

**Figure 5**. (a), (b) ADF STEM images of the same Fe-N nanocrystal and nitrogen inclusion as in Fig. 4, recorded while drilling a hole in the specimen using a stationary 100 kV focused electron beam after (a) 0 and (b) 640 seconds. (c) – (e). Background-subtracted EELS spectra taken from a time series of N-*K* edge measurements. After approximately 10 min. (between spectra (d) and (e)), a hole forms in the specimen and the nitrogen is released from the inclusion. The intensity of the first peak in the spectrum is then reduced.

**IV. DISCUSSION**

The study of a molecular-nitrogen filled inclusion in solid GaN using TEM and EELS is very challenging. First, the specimen undergoes radiation damage by the electron beam, including (i) ionization (radiolysis), (ii) sputtering by knock-on and (iii) specimen heating.[30,31] Ionization is likely to result in a chemical shift of the N-*K* edge, but not to have a significant effect on the overall intensity of the EELS spectrum. Knock-on damage of GaN is also unlikely, since the bulk threshold knock-on energies for N and Ga atom displacements are 32 and 24 eV, which require electron energies of 180 and 510 keV, respectively, for the production of Frenkel pair point defects.[32] With regard to the nitrogen in the inclusion, the combined effect of displacement and ionization can result in the weakening or splitting of atomic bonds in the nitrogen dimers. The complexity of the system is potentially even greater as a result of the presence of Fe in the vicinity of the inclusion, since an Fe-based catalyst is used for splitting nitrogen bonds in the presence of hydrogen in the Haber-Bosch process.[33] With regard to specimen heating, the temperature rise[31] of the specimen is expected to be given by the expression $\Delta T \sim \langle E \rangle (2R_0/b) / (4\pi\kappa\lambda)$, where $\langle E \rangle$ is the mean energy loss per inelastic scattering event, $R_0$ is the distance from the beam position to the conductive part of the TEM stage or grid bar, $b$ is the probe size, $\kappa$ is the thermal conductivity of the specimen and $\lambda$ is the inelastic mean free path. A 100 kV STEM probe is therefore expected to increase the temperature of a ~100 nm thick specimen by only a few degrees, as GaN has a thermal conductivity of $\kappa$ = 130 W m$^{-1}$K$^{-1}$. However, the molecular nitrogen gas has a thermal conductivity of $\kappa$ = 0.026 W m$^{-1}$K$^{-1}$, which is four orders of magnitude lower than that of GaN. Electron-beam-induced heating may therefore be negligible for GaN at 100 kV, but it is less well understood for nitrogen gas in GaN. Additional energy cascade processes, e.g. involving photoelectrons and Auger electrons, may also transfer energy to the GaN host rather than to the nitrogen gas, due to the greater mean free path of electrons in the gas than in the inclusion.

The complexity of the experiment performed on the nitrogen-containing inclusion in GaN is also illustrated by the dynamic transformation of the inclusion shape during STEM imaging and EELS, as shown in Fig. 6. The truncated shape of the inclusion is seen to transform first into a trapezoid and then to a triangular shape, thereby reducing its contact area with the Fe-N nanocrystal, as shown in Figs. 6 (a) – (d). The size of the Fe-N nanocrystal does not change significantly. Only the interface between the nanocrystal and the inclusion becomes more curved during the experiment, as marked by arrows in Figs. 6 (b) – (d).

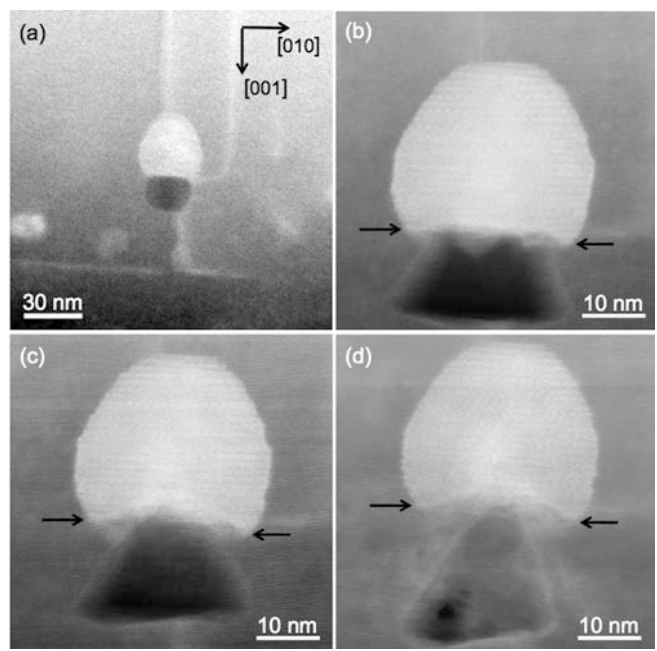

**Figure 6.** ADF STEM images of the same Fe-N and nitrogen inclusion complex as in Figs. 4 and 5, showing shape changes undergone by the nanocrystal and the inclusion during the experiments. (a) One of the first scans performed at low magnification; (b) a few scans later at medium resolution; (c) ~ 14 minutes later and (d) after the hole-drilling experiment, ~ 120 minutes after (b). The black arrows indicate changes to the interface between the Fe-N nanocrystal and the nitrogen inclusion. The experimental conditions are the same as in Fig. 4.

The nitrogen pressure in the inclusion can in principle be determined from an EELS measurement by using the expression $[I_{i+GaN}/I_{GaN}] = [\rho_{N(i)} \cdot d_i + \rho_{GaN} \cdot (d_{GaN} - d_i)]/\rho_{GaN} \cdot d_{GaN}$ where $I$, $\rho$, and $d$ are the integrated intensities of the energy-loss peaks, the nitrogen densities and the specimen thicknesses, respectively, of the nitrogen inclusion and the GaN host. The total specimen thickness was measured to be $d_{GaN+i}=110 \pm 10$ nm from a low-loss EELS intensity measurement. There are 44 nitrogen atoms per nm$^3$ in GaN, which corresponds to a nitrogen density $\rho_{N(GaN)}$ of 1.026 g/cm$^3$. The background-subtracted N-$K$ edge peaks were integrated in the energy range between 398 and 448 eV. On the assumption of single scattering and that the inclusion is spherical with a diameter $d_i \sim 20$ nm, the density of nitrogen was estimated to be $1.4 \pm 0.3$ g/cm$^3$, which corresponds to a gas pressure of ~ 3 GPa at 300 K according to a N$_2$ pressure – density isotherm calculated by Strak et al.,[34] or to ~ 2.8 GPa according to a volume – pressure diagram for N$_2$ determined by Mills et al.[35] Interestingly, the measured density is similar to that of solid nitrogen.[35] At the same time, the EELS spectrum suggests that it is probably in a molecular state, while the lack of Moiré patterns in STEM images suggests that it is amorphous. Inclusions with higher densities, containing (probably solid) nitrogen, have been found in sapphire close to a GaN/sapphire interface by Matsubara et al.,[36] due to nitridation of the surface. Our pressure estimate is simplified, as it does not consider differences in scattering cross-section between the inclusion and the GaN or the unknown temperature of the inclusion resulting from the large thermal conductivity difference between nitrogen gas and the GaN host. An approach similar to that used by Walsh *et al.*[37] to measure the density and pressure of gas in a helium bubble in an irradiated Ni-Fe-Cr alloy could be used in a future study of the pressure of nitrogen-filled inclusion adjacent to Fe-N nanocrystals in GaN.

**IV. SUMMARY**

In summary, nitrogen-filled inclusion adjacent to Fe$_n$N ($n$ = 3 or 4) nanocrystals in (Ga,Fe)N have been identified and studied using aberration-corrected ADF STEM and EELS. The Fe$_n$N nanocrystals are arranged in a planar array in the GaN matrix. Typically, nanocrystals that are larger than ~ 5 nm are found to be associated with nitrogen-containing inclusion in samples deposited above 800 °C. Larger Fe$_n$N nanocrystals appear to be truncated at their boundaries with the adjacent to such inclusion. ADF STEM images recorded as a function of camera length suggest the presence of strain in GaN around the nitrogen-filled inclusion. The nitrogen density in an inclusion formed in a sample deposited at 850 °C is estimated to be ~1.4 g/cm$^3$. The nitrogen inclusion shows strong shape transformations under electron beam illumination. An *in situ* hole drilling experiment is used to record N-*K* edge spectra before and after the nitrogen in the inclusion is released.

The presence of nitrogen inclusion provides an explanation for the location of surplus nitrogen, which is liberated by the nucleation of Fe$_n$N (n>1) nanocrystals during the growth of (Ga,Fe)N epilayers. As shown in reference [4], optimization of the growth parameters during the deposition of (Ga,Fe)N can be used to control the aggregation and structure of the Fe$_n$N inclusions and, in principle, to eliminate them.

**Acknowledgement**

The work was supported by the European Research Council through the FunDMS Advanced Grant (#227690) within the "Ideas" 7th Framework Programme of the EC, and by the Austrian FondszurFörderung der wissenschaftlichen Forschung – FWF (P18942, P20065 and N107-NAN). The EPSRC is acknowledged for financial support under grant reference EP/D040205/1. M.S.M. acknowledge partial financial support from CONICET (Argentina). D.A. Muller and M. Somers are gratefully acknowledged for valuable discussions.

**References**

1. A. Bonanni, and T. Dietl, Chem. Soc. Rev. 39, 528 (2010).

2. K. Sato, L. Bergqvist, J. Kudrnovský, P. H. Dederichs, O. Eriksson, I. Turek, B. Sanyal, G. Bouzerar, H. Katayama-Yoshida, V. A. Dinh, T. Fukushima, H. Kizaki, and R. Zeller, Rev. Mod. Phys. 82, 1633 (2010).

3. N. Gonzalez Szwacki, J. A. Majewski, and T. Dietl, Phys. Rev. B 83, 184417 (2011).

4. A. Navarro-Quezada, T. Devillers, T. Li, and A. Bonanni, Appl. Phys. Lett. 101, 081911 (2012)

5. A. Bonanni, A. Navarro-Quezada, T. Li, M. Wegscheider, Z. Matej, V. Holy, R. T. Lechner, G. Bauer, M. Rovezzi, F. D'Acapito, M. Kiecana, M. Sawicki, and T. Dietl, Phys. Rev. Lett. 101, 135502 (2008).

6    A. Navarro-Quezada, N. Gonzalez Szwacki, W. Stefanowicz, T. Li, A. Grois, T. Devillers, M. Rovezzi, R. Jakieła, B. Faina, J. A. Majewski, M. Sawicki, T. Dietl, and A. Bonanni, Phys. Rev. B 84, 155321 (2011).

7    I. A. Kowalik, A. Persson, M. A. Nino, A. Navarro-Quezada, B. Faina, A. Bonanni, T. Dietl, and D. Arvanitis, Phys. Rev. B 85, 184411 (2012).

8    M. Birowska, C. Śliwa, J. A. Majewski, and T. Dietl, Phys. Rev. Lett. (2012).

9	J. De Boeck, R. Oesterholt, A. Van Esch, H. Bender, C. Bruynseraede, C. Van Hoof, and G. Borghs, Appl. Phys. Lett. 68, 2744 (1996).

10.	M. Moreno, A. Trampert, B. Jenichen, L. Däweritz, and K. H. Ploog, J. Appl. Phys. 92, 4672 (2002).

11.	M. Yokoyama, H. Yamaguchi, T. Ogawa, and M. Tanaka, J. Appl. Phys. 97, 10D317 (2005).

12.	J. Sadowski, J. Z. Domagała, R. Mathieu, A. Kovács, T. Kasama, R. E. Dunin-Borkowski, and T. Dietl, Phys. Rev. B 84, 245306 (2011).

13.	S. Hara, M. Lampalzer, T. Torunski, K. Volz, W. Treutmann, and W. Stolz, J. Cryst. Growth 261, 330 (2004).

14.	J. H. Park, M. G. Kim, H. M. Jang, S. Ryu, and Y. M. Kim, Appl. Phys. Lett.84, 1338 (2004).

15.	R. Lardé, E. Talbot, P. Pareige, H. Bieber,G. Schmerber, S. Colis, V. Pierron-Bohnes, and A. Dinia, J. Am. Chem. Soc. 133, 1451 (2011).

16.	M. Sawicki, E. Guziewicz, M. I. Łukasiewicz, O. Proselkov, I. A. Kowalik, W. Lisowski, P. Dłużewski, A. Wittlin, M. Jaworski, A. Wolska, W. Paszkowicz, R. Jakieła, B. S. Witkowski, L. Wachnicki, M. T. Klepka, F. J. Luque, D. Arvanitis, J. W. Sobczak, M. Krawczyk, A. Jablonski, W. Stefanowicz, D. Sztenkiel, M. Godlewski, and T. Dietl, arXiv:1201.5268.

17.	T. Dietl, Nat. Mater. 9, 965 (2010).

18.	J. M. D. Coey, K. Wongsaprom, J. Alaria, and M. Venkatesan, J. Phys. D: Appl. Phys. 41, 134012 (2008).

19.	Y. Yamada, K. Ueno, T. Fukumura, H. T. Yuan, H. Shimotani, Y. Iwasa, L. Gu, S. Tsukimoto, Y. Ikuhara, and M. Kawasaki, Science 332, 1065 (2011).

20. W. Heimbrodt, P. J. Klar, S. Ye, M. Lampalzer, C. Michel, S. D. Baranovskii, P. Thomas, and W. Stolz, J. Supercond.. 18, 315 (2005).

21. R. Akiyama, S. Ohya, P. N. Hai, and M. Tanaka, J. Appl. Phys. 111, 063716 (2012).

22. H. Katayama-Yoshida, K. Sato, T. Fukushima, M. Toyoda, H. Kizaki, V. A. Dinh, and P. H. Dederichs, phys. stat. sol. (a) 204, 15 (2007).

23. T. Dietl, J. Appl. Phys. 103, 07D111 (2008).

24. A. Navarro-Quezada, W. Stefanowicz, T. Li, B. Faina, M. Rovezzi, R. T. Lechner, T. Devillers, F. d'Acapito, G. Bauer, M. Sawicki, T. Dietl, and A. Bonanni, Phys. Rev. B 81, 205206 (2010).

25. M. S. Moreno, S. Lazar, H.W. Zandbergen and R.F. Egerton, Phys. Rev. B 73, 073308 (2006)

26. Z. Yu, D.A. Muller, and J. Silcox, J. Appl. Phys. 95, 3362 (2004).

27. A. Leineweber, H. Jacobs, F. Hüning, H. Lueken, and W. Kockelmann, J. All. Comp. 316, 21 (2001)

28. K. Sader, B. Schaffer, G. Vaughan, R. Brydson, A. Brown, and A. Bleloch, Ultramicroscopy 110, 998 (2010).

29. J. R. Jinschek and S. Helveg, Micron 43, 1156 (2012)

30. D. B. Williams and C. B. Carter, Transmission Electron Microscopy: A Textbook for Materials Science, (Springer 2009) pp. 64-68

31. R.F. Egerton, Ultramicroscopy 127, 100 (2012)

32. A. Ionascut-Nedelcescu, C. Carlone, A. Houdayer, H. J. von Bardeleben, J.-L. Cantin, and S. Raymond, IEEE Trans. Nucl. Sci. 49, 2733 (2002)

33. See, for example, M. D. Fryzuk and S.A. Johnson, Coor. Chem. Rev. 200-202, 379 (2000)

34. P. Strak and S. Krukowski, J. Chem. Phys. 126, 194501 (2007)

35. R. L. Mills, B. Olinger, and D. T. Cromer, J. Chem. Phys. 84, 2837 (1986)

36. T. Matsubara, and K. Shoda, Jap. J. Appl. Phys. 45, 279 (2006)

37. C. A. Walsh, J. Yuan, and L.M. Brown, Phil. Mag. A 80, 1507 (2000)